# Role of temperature-dependent electron trapping dynamics in the optically driven nanodomain transformation in a PbTiO$_3$/SrTiO$_3$ superlattice

Joonkyu Park,[1] Youngjun Ahn,[1] Jack A. Tilka,[1,*] Hyeon Jun Lee,[1] Anastasios Pateras,[1] Mohammed H. Yusuf,[2,*] Matthew Dawber,[2] Haidan Wen,[3] and Paul G. Evans[1,†]

[1] *Department of Materials Science and Engineering, University of Wisconsin-Madison, Madison, WI 53706, USA*

[2] *Department of Physics and Astronomy, Stony Brook University, Stony Brook, NY 11794, USA*

[3] *Advanced Photon Source, Argonne National Laboratory, Argonne, Illinois 60439, USA*

The spontaneously formed striped polarization nanodomain configuration of a PbTiO$_3$/SrTiO$_3$ superlattice transforms to a uniform polarization state under above-bandgap illumination with a time dependence varying with the intensity of optical illumination and a well-defined threshold intensity. Recovery after the end of illumination occurs over a temperature-dependent period of tens of seconds at room temperature and shorter times at elevated temperatures. A model in which the screening of the depolarization field depends on the population of trapped electrons correctly predicts the observed temperature and optical intensity dependence.

Epitaxial ferroelectric heterostructures and superlattices exhibit a range of complex polarization configurations, including nanodomain stripes, ferroelectric vortices, and skyrmions and raise the possibility that these configurations can be manipulated through nanoscale patterning.[1-5] Polarization configurations depends sensitively on the screening of the polarization by free charges and charge at surfaces and interfaces, which can be varied by creating a high population of optically excited charge carriers or via subsequent optically induced changes in the population of traps.[6-8] Optically induced carrier dynamics and charge trapping effects expand the range of phenomena available for the development of nanoscale polarization configurations. Precise optical control of nanoscale polarization also has the potential to yield optically switchable devices and electronic materials.

The ferroelectric polarization of $PbTiO_3/SrTiO_3$ (PTO/STO) ferroelectric/dielectric superlattices (SLs) spontaneously forms an intricate striped polarization pattern with nanometer-scale periodicity.[9,10] The lateral period is on the order of 10 nm in SLs with few-nm repeating unit thicknesses.[11] Above-bandgap optical illumination induces a transformation in which the striped nanodomain pattern of these SLs changes to a uniform polarization state through a mechanism linked to depolarization field screening.[12] Similar optically driven transformations are also observed in other SLs with the same composition but more complex initial polarization states.[13]

Ultrafast above-bandgap optical excitation of ferroelectrics leads to a transient lattice expansion driven by excited charge carriers.[14-17] In $BiFeO_3$ thin films, for example, the expansion arises on the picosecond timescale of acoustic pulse propagation through the layer thickness and subsequently decays as carriers recombine.[14,16] In comparison with $BiFeO_3$, however, PTO/STO SLs exhibit a complex recovery over a period of seconds at room temperature, which is not yet clearly explained.[12]

This Letter reports time-resolved synchrotron x-ray diffraction measurements showing that the rate of the domain transformation depends strongly on the temperature and optical intensity. The time required for the transformation to the uniform polarization state, for example, is reduced by a factor of 5 when the optical intensity is increased from 1.1 W cm$^{-2}$ to 2.8 W cm$^{-2}$. The recovery after the end of the illumination becomes dramatically faster at elevated temperatures. The rates of the transformation and subsequent recovery are accurately predicted by a model of the population of trapped charges and associated screening of the depolarization field.

X-ray diffraction studies were conducted at station 7-ID-C of the Advanced Photon Source of Argonne National Laboratory using the experimental arrangement in Fig. 1(a). An x-ray beam with 11 keV photon energy was focused to a 500 nm full width at half maximum (FWHM) spot on the SL. The diffracted x-ray intensity was recorded by a pixel array detector (Pilatus 100 K, Dectris, Ltd.). The pulsed optical pump beam had 60-fs duration, a wavelength of 400 nm (photon energy 3.1 eV), repetition rate of 1 kHz, and spot size 140 μm FWHM. With the sample is oriented to meet the Bragg condition for the 002 reflection of the SL, the optical beam was incident at an angle of 16° with respect to the surface normal. A comparison of the domain diffuse scattering patterns acquired with this arrangement and patterns acquired using less-intense laboratory x-ray sources indicates that the incident x-ray beam does not perturb the steady-state domain diffuse x-ray scattering. It is in principle possible, however, that the x-rays have a subtle effect on the overall recovery dynamics. There is no evidence from the experimental data here, however, for that possibility. The optical absorption length in the PTO/STO SL is 1 μm for a photon energy of 3.1 eV, calculated using an effective medium approximation using the optical constants of PTO and STO.[18-21] The PTO/STO SL thin film heterostructure was thus uniformly optically excited.

The SL consisted of 23 repeats of 8 unit cells (u.c.) of PTO and 3 u.c. of STO, deposited on an SrRuO$_3$ (SRO) thin film on an STO (001) substrate using off-axis radio-frequency sputtering. The 002 SL x-ray reflection is at out-of-plane wavevector $Q_z$ = 3.13 Å$^{-1}$ with a corresponding ring of domain diffuse scattering intensity with in-plane radius 0.072 Å$^{-1}$ at the same $Q_z$, as illustrated Fig. 1(b). The observation of a ring of diffuse scattering indicates that the ferroelectric striped nanodomains have random in-plane orientation. The in-plane reciprocal-space width of the domain diffuse scattering maximum corresponds to an in-plane stripe pattern coherence length of a few periods. The average SL lattice parameter in the out-of-plane direction, $c_{SL}$, and domain period $\Lambda$ were 4.016 Å and 8.7 nm, respectively.

The domain diffuse scattering disappears during illumination, as in the lower panel of Fig. 1(b). The transformation is accompanied by a shift of the structural x-ray reflection to lower wavevector $Q_z$ by $\Delta Q_z$. Together, these observations are consistent with an optically induced domain transformation to a uniform polarization state. It is in principle possible that the final state is a more complex polarization configuration, but our experiments have not revealed any new diffuse scattering features after the transformation. The out-of-plane reciprocal-space width of the domain diffuse scattering does not vary during the disappearance or recovery of the domain pattern, which indicates that coherence length of the domain pattern along the surface normal is unchanged during these processes. The absolute value of the out-of-plane coherence length is approximately equal to the total SL thickness at all times, indicating that all layers of the SL are transformed together in each location.

The nanodomain recovery after the end of illumination leads to a reappearance of the domain diffuse scattering and a return of the lattice parameter to its initial value at a rate that varies as a function of temperature. Figs. 1(c) and 1(d) show the reciprocal-space distribution of domain diffuse scattering intensity at two temperatures before illumination, at 10 s after the start of illumination 1.1 W cm$^{-2}$, and at 5 s after the end of the illumination. At both room temperature, Fig. 1(c), and 335 K, Fig. 1(d), the diffuse scattering signals disappear during illumination. At room temperature, the domain diffuse scattering intensity remains low at 5 s after the end of illumination. At 335 K, however, the domain diffuse scattering has nearly fully recovered by this time.

The time dependence of the wavevector of the SL Bragg reflection and the intensity of the nanodomain diffuse scattering after optical illumination at 1.1 W cm$^{-2}$ are shown for several temperatures in Fig. 2. The wavevector of the SL Bragg reflection was measured by fitting the diffraction pattern acquired at each time with a Gaussian peak. The domain intensities shown in Fig. 2(b) are normalized with respect to the domain diffuse scattering intensity before optical excitation. The time required to recover 90% of the initial domain diffuse scattering intensity is plotted as a function of temperature in Fig. 2(c). The 90% recovery time is 120 s at room temperature and decreases to 5 s at 335 K.

The optically induced transformation from the nanodomain state to the uniform polarization configuration can be described using a model in which the depolarization field driving the formation of domains is screened by optically excited carriers.[12] The screening is parameterized in a Landau-Ginsburg-Devonshire model of the transformation using a screening coefficient $\theta$ that ranges from 0 (unscreened) to 1 (fully screened).[12] The free energy densities calculated for the uniform polarization and nanodomain configurations are plotted in Fig. 3(a) as function of $\theta$. The nanodomain state is energetically favored when the depolarization field is unscreened at $\theta=0$. As $\theta$ increases, the total free energies for both cases change due to the depolarization field screening effect. For values of $\theta$ greater than 0.78, the uniform polarization state has lower free energy than the nanodomain configuration. A similar threshold phenomenon is observed at a slightly higher value of the screening parameter in computational studies of individual ultrathin PTO and BaTiO$_3$ layers.[22] We define the threshold charge concentration required to reach this critical value of $\theta$ to be $N_{th}$.

Optical excitation leads to a change in carrier concentration and thus to a change in the screening. The recombination time constant for optically excited carriers has been previously observed to be on the order of nanoseconds to microseconds in ferroelectric thin films, far shorter than the recovery timescale evident in Fig. 1.[23,24] A significant population of carriers, however, can be trapped at defects before the fast recombination.[25] Electron and hole traps in ferroelectrics arise from valences changes of metal ions,[26] metal or oxygen vacancies or vacancy complexes,[27,28] and impurities.[29] Trap energies vary widely, but generally range from 0.5 eV to 0.8 eV, and trapped charges can be trapped for durations on the order of seconds.[25,28,30] With long-lived traps, the number of accumulated charge carriers depends on the total illumination time. With a sufficiently high optical intensity and long trap lifetimes the illumination can lead to a carrier concentration above $N_{th}$, inducing the domain transformation. After the end of illumination, the accumulation of trapped charges ceases and de-trapping leads to a decrease of the trapped charge concentration and the recovery of the nanodomain pattern when the concentration falls below $N_{th}$.

The time dependence of the recovery of the domain diffuse scattering intensity after the end of illumination suggests that the transformation proceeds heterogeneously, with varying threshold carrier concentrations, trap time constants, or trap concentrations in different spatially separated regions. The experimentally observed overall return of the domain diffuse scattering intensity is non-exponential and we thus consider here a set of traps with different energies $E_{T,i}$ indexed by the integer $i$. Carriers leave the traps and recombine with a time constant $\tau_i(T) = \frac{1}{C \exp\left(-\frac{E_{T,i}}{k_B T}\right)}$.[31] Here $k_B$ is the Boltzmann constant, $C$ is a recombination rate constant that we assume to be the same for all traps, and $T$ is the temperature.

The concentration of trapped charges increases immediately following each optical pulse and relaxes during the interval between them. The trapped charge concentration resulting from a single optical pulse is $N_t$. The probability that a trap state $i$ that is initially occupied at the time of optical excitation remains occupied after a time equal to the optical repetition rate is $P_{d,i} = exp\ (-\frac{t_r}{\tau_i(T)})$. The interval between optical pulses was $t_r = 1$ ms. From a sum of the corresponding geometric series, the population $N_{accum,i}(t_e,T)$ in the traps with index $i$ at elapsed time $t_e$ during illumination is:

$$N_{accum,i}(t_e, T) = N_t P_{d,i} \left[\frac{P_{d,i}^{(\frac{t_e}{t_r})} - 1}{P_{d,i} - 1}\right].$$

The domain transformation occurs when $N_{accum,i}$ exceeds $N_{th}$, as illustrated in Fig. 3(b). After illumination ends the concentration of charges in each trap decreases exponentially with time constant $\tau_i$.

The domain diffuse scattering intensity can be predicted by assuming that the relative contribution to domain diffuse scattering intensity from the region with index $i$ is given by the unit step function $\mathrm{H}(N_{accum,i}(t_e, T) - N_{th})$. The total domain diffuse scattering intensity at elapsed time $t_e$ during illumination is:

$$I(t_e) = \frac{I_0}{n} \sum_{i=1}^{n} \int_0^{t_e} \frac{1 - \mathrm{H}(N_{accum,i}(t', T) - N_{th})}{2} dt'$$

Here $I_0$ is the initial domain intensity and $n$ is the number of distinct trap energies. The simulation presented here includes a large enough value of $n$ to approximate a continuous distribution of trap energies with equal concentration per unit energy.[25,30] The precise value of $n$, and the range of energies for which agreement between the experiment and model was achieved, both varied slightly from location to location on the sample surface. The model in Fig. 3c used *n=26* trap energies over the range from 800 to 900 meV. The model in Fig. 4 used *n=21* trap energies in a range from 820 to 900 meV.

The predicted time dependence of domain diffuse scattering intensity at room temperature and 335 K is shown in Fig. 3(c), which also compares the predicted intensity with experimental observations from Fig. 2(b). The trapping model reproduces the time dependence of the recovery at both temperatures, with 90% recovery times of 80 s at room temperature and 1.5 s at 335 K.

The charge trapping model also predicts the dependence of the domain transformation on the optical intensity. The predicted time dependence of $N_{accum}/N_{th}$ for optical intensities of 1.1 and 2.8 W cm$^{-2}$ and a single trap energy level at 860 meV are shown in Fig. 3(d). The domain transformation occurs faster at higher optical intensity, with $N_{accum}$ exceeding $N_{th}$ at 8 s and 2 s after the start of illumination for 1.1 and 2.8 W cm$^{-2}$, respectively. $N_{accum}$ decreases immediately after the end of illumination and passes $N_{th}$ at 13 s and 37 s at 1.1 and 2.8 W cm$^{-2}$.

Figure 4(a) shows the time dependence of $\Delta Q_z$ for optical intensities of 0.9, 1.1 and 2.8 W cm$^{-2}$. At room temperature, $\Delta Q_z$ increases under illumination and relaxes over a period of seconds after the end of illumination. The values of $\Delta Q_z$ are small at 0.9 W cm$^{-2}$ because the charge recombination is faster than accumulation at this optical intensity and the trapped charge does not reach the threshold. At 1.1 W cm$^{-2}$, however, $\Delta Q_z$ clearly increases as a function of time and reaches 0.02 Å$^{-1}$ during the illumination period. At 2.8 W cm$^{-2}$, $\Delta Q_z$ reaches and saturates at 0.03 Å$^{-1}$ within a few seconds. At $T = 400$ K and 2.8 W cm$^{-2}$, there are no apparent changes in $\Delta Q_z$ because the recombination rate is significantly higher at elevated temperature.

Figure 4(b) shows the time dependence of the domain diffuse scattering intensity during and after illumination. The fractional decrease in domain scattering intensity after 25 s at optical intensity 0.9 W cm$^{-2}$ is 12.5% and the transformation to the uniform polarization state never occurs completely. At higher optical intensity the transformation occurs with 90% completion times of 11.5 s at 1.1 W cm$^{-2}$ and 2 s at 2.8 W cm$^{-2}$. The 90% completion times for the domain transformation at room temperature are plotted as a function of optical intensity in Fig. 4(c).

The predicted and observed time dependence of the domain diffuse scattering intensity are compared in Fig. 4(b) using the same model parameters as above. The prediction reproduces the dependence of the rate of the domain transformation on the optical intensity and the recovery after the end of illumination. The recovery dynamics are similar at both optical intensities because the recovery occurs in the absence of optical illumination and thus depends only on the initial trap populations.

Time-resolved x-ray diffraction studies reveal that optically drive destabilization of the nanodomain pattern of PTO/STO SLs is driven by the carrier population within traps. As was the case, with the pinning of domain walls by trapped photoinduced charges at far larger lengthscales, the dynamics are consistent with thermally activated transitions out of trap states.[29] The dynamics of the transformation are faster at higher optical intensity due to faster charge accumulation. An optimization of optical intensity and the temperature can in principle facilitate the ultrafast optical control of the nanodomains and other polarization features.

X-ray scattering studies were supported by the U.S. Department of Energy (DOE) Office of Science Basic Energy Sciences through Award No. DE-FG02-04ER46147 (P.E.). Thin film synthesis efforts were supported by the U.S. NSF through Grants No. DMR-1055413 and No. DMR-1334867 (M.D.). This research used resources of the Advanced Photon Source, a U.S. DOE Office of Science User Facility operated for the DOE Office of Science by Argonne National Laboratory under Contract No. DE-AC02-06CH11357. The authors gratefully acknowledge use of facilities and instrumentation supported by NSF through the University of Wisconsin Materials Research Science and Engineering Center (DMR-1720415).

*Present address: Intel Corporation, Hillsboro, Oregon 97124, USA

†pgevans@wisc.edu

**Figure 1.** (a) Optically induced domain transformation in a PTO/STO SL. (b) Reciprocal-space locations of the SL 002 reflection (red) and domain diffuse scattering (green ring). After the transformation, the SL reflection shifts by $\Delta Q_z$, and the domain intensity disappears. (c) and (d) Domain diffuse scattering intensity at room temperature and 335 K before illumination, during illumination following the transformation to the uniform polarization state, and 5 s after the end of illumination.

**Figure 2.** (a) $\Delta Q_z$ and (b) domain diffuse scattering intensity as a function of time after the end of illumination at room temperature, 310 K, and 335 K with optical intensity 1.1 W cm$^{-2}$. (c) Temperature dependence of the 90% recovery time of the domain intensity.

**Figure 3.** (a) Free energy density of uniform-polarization and nanodomain states as a function of screening coefficient for a Landau-Ginsburg-Devonshire calculation. The threshold value of the screening, above which the uniform polarization is stable, has trap population $N_{th}$. (b) Dynamics of trap population in the period of several optical pulses. (c) Predicted and observed time dependence of domain diffuse scattering recovery at room temperature and 335 K. (d) Predicted values of $N_{accum}/N_{th}$ as a function of time during and after the end of illumination at optical intensities 1.1 and 2.8 W cm$^{-2}$. The shaded region at negative times represents the 25 s duration of optical illumination.

**Figure 4.** (a) Observed shift $\Delta Q_z$ and (b) observed (points) and predicted (solid lines) domain diffuse scattering intensity as a function of time during and after illumination for optical intensities of 0.9, 1.1, and 2.8 W cm$^{-2}$ at room temperature (upper panels) and 400 K (bottom

panels). (c) Time for 90% reduction of domain diffuse scattering intensity as a function of optical intensity.

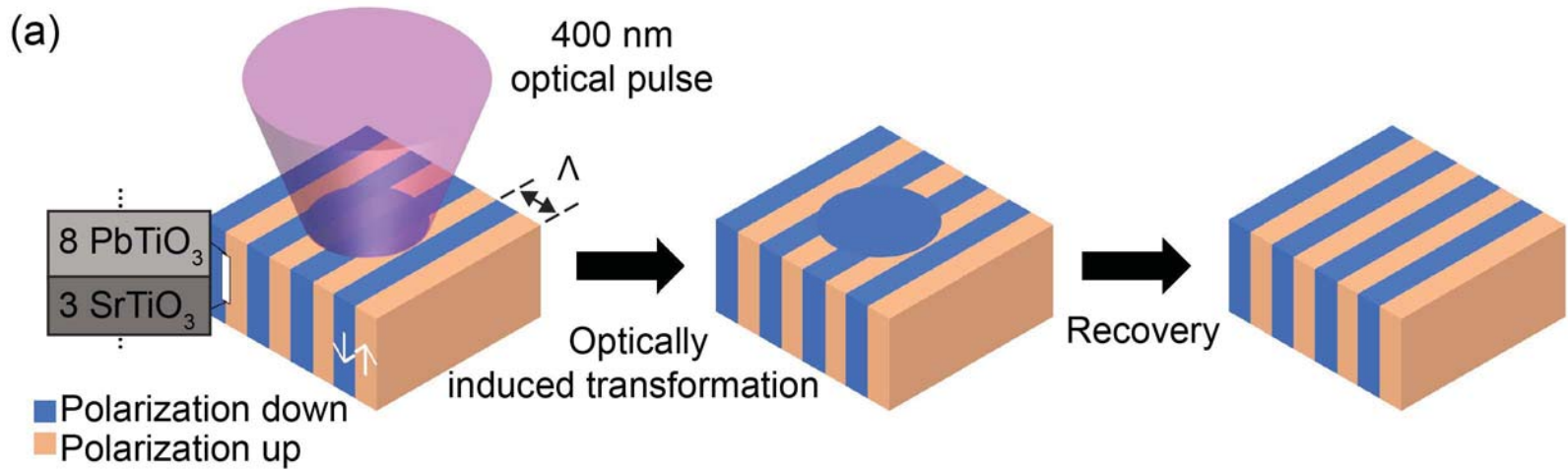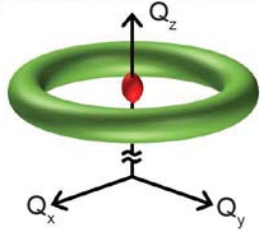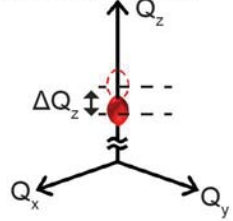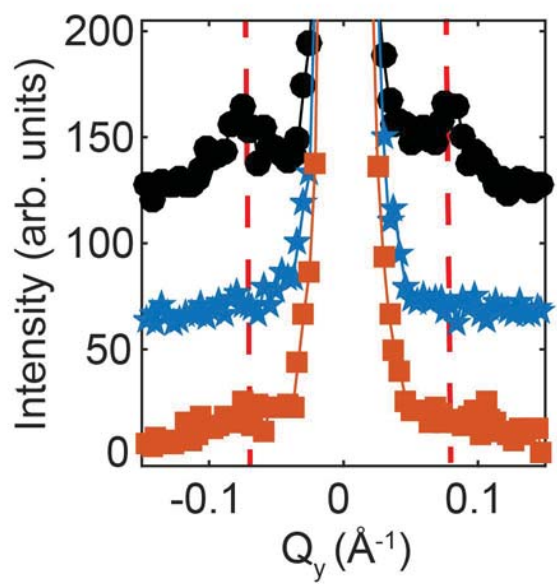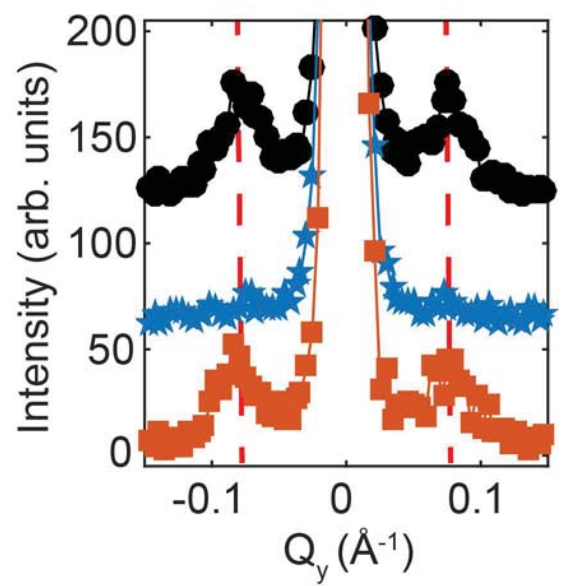

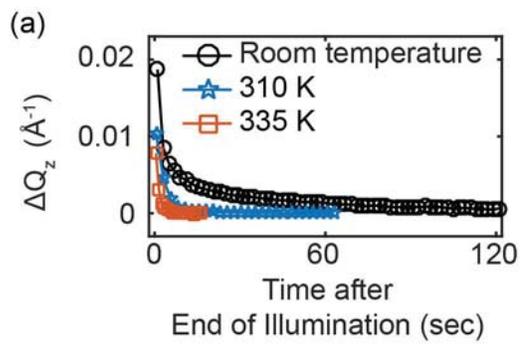

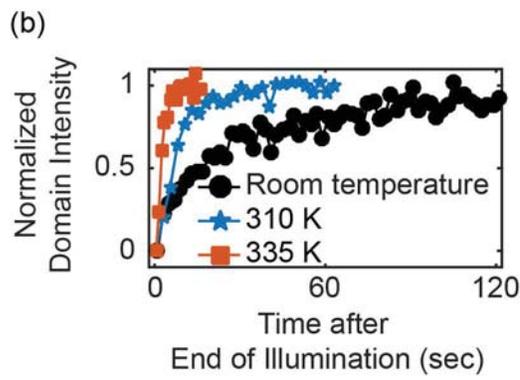

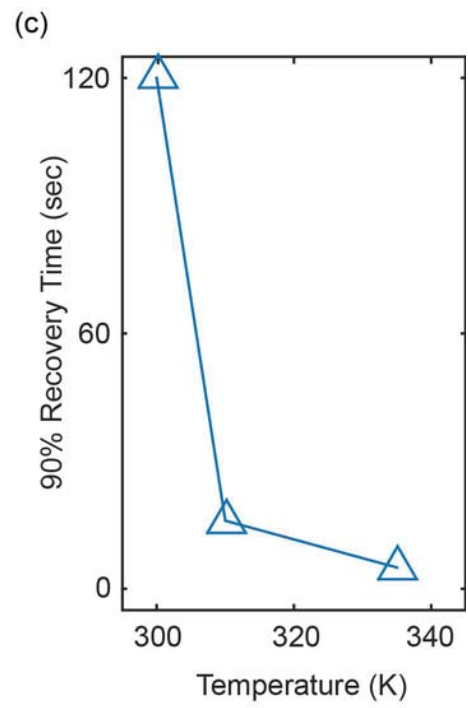

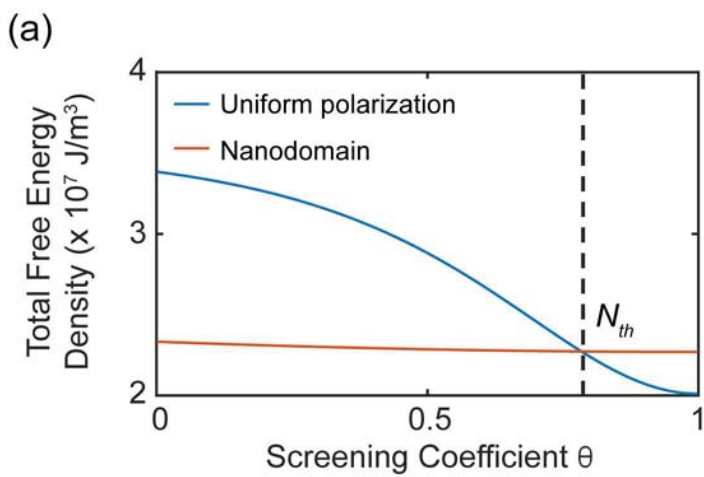
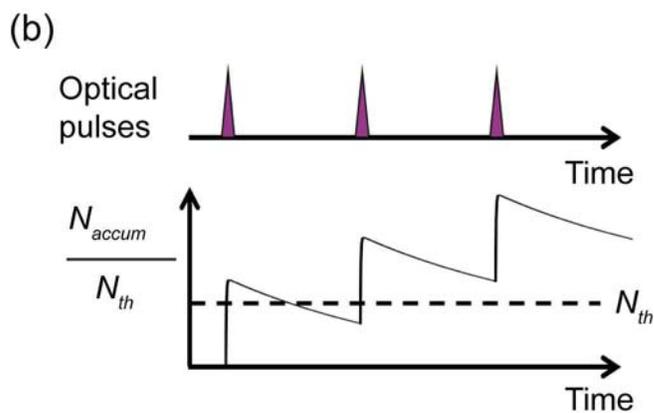
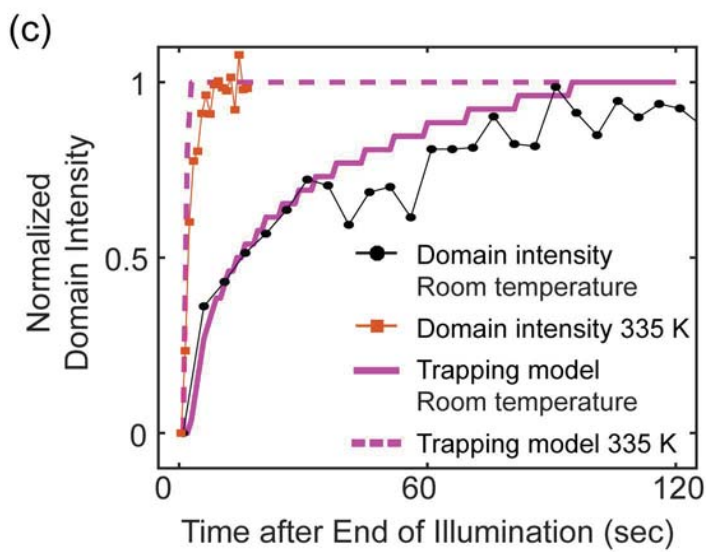
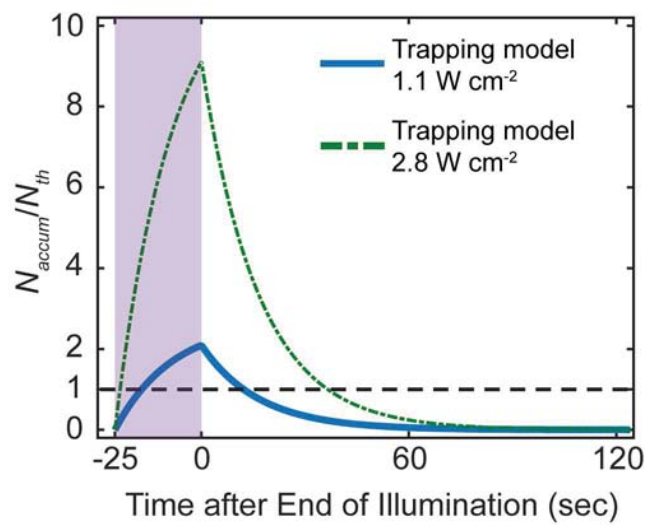

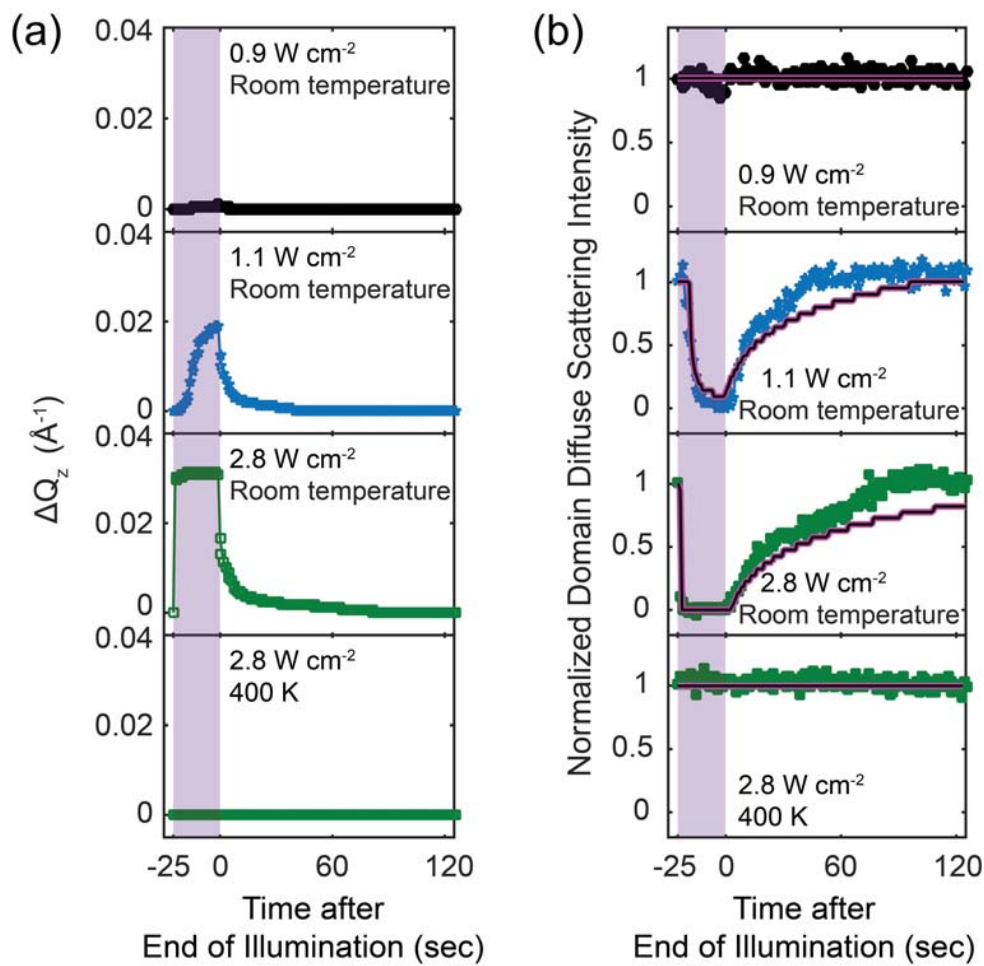

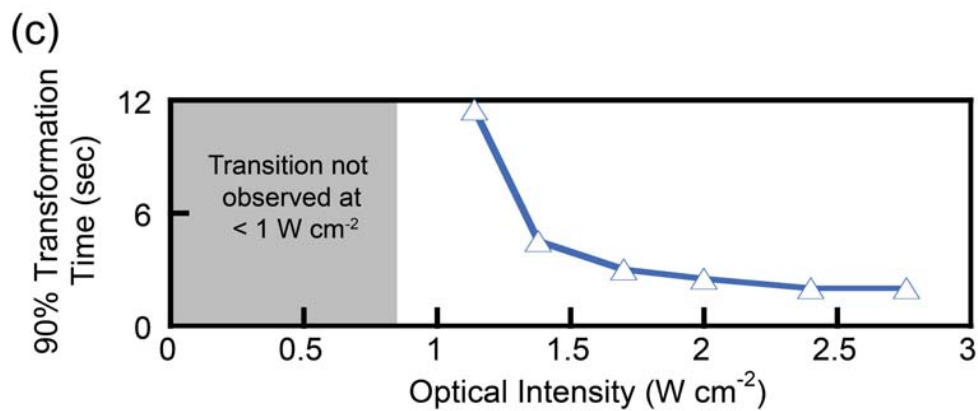